 \documentclass[10pt,conference]{IEEEtran}  

\usepackage{latexsym}
\usepackage{graphicx}
\usepackage{amsmath}
\usepackage{amssymb}
\usepackage{psfrag}
\usepackage{color}
\usepackage{cite}
\usepackage{times}
\usepackage{setspace}
\usepackage{multirow}

\newcommand{\stxt}[1]{\ensuremath{_{\text{#1}}}}

\long\def\symbolfootnote[#1]#2{\begingroup%
\def\thefootnote{\fnsymbol{footnote}}\footnote[#1]{#2}\endgroup}
\title{\LARGE \bf
Connecting Spatially Coupled LDPC Code Chains}

\author{
\authorblockN{Dmitri Truhachev$^*$, David G. M. Mitchell$^\dag$, Michael Lentmaier$^\ddag$, and Daniel J. Costello, Jr.$^\dag$}
\authorblockA{$^*$Department of Computing Science, University of Alberta, Edmonton, Canada\\dmitryt@ualberta.ca\\
$^\dag$Dept. of Electrical Engineering, University of Notre Dame, Notre Dame,
Indiana, USA,\\
\{david.mitchell, costello.2\}@nd.edu\\
$^\ddag$Vodafone Chair Mobile Communications Systems,
       Dresden University of Technology, Dresden, Germany,\\
michael.lentmaier@ifn.et.tu-dresden.de}\vspace{-8mm}}

\begin{document}

\maketitle

\begin{abstract}
Codes constructed from connected spatially coupled low-density parity-check code (SC-LDPCC) chains are proposed and analyzed. It is demonstrated that connecting coupled chains results in improved iterative decoding performance. The constructed protograph ensembles have better iterative decoding thresholds compared to an individual SC-LDPCC chain and require less computational complexity per bit when operating in the near-threshold region. In addition, it is shown that the proposed constructions are asymptotically good in terms of minimum distance. 
\end{abstract}

\section{Introduction}


Recently, iterative processing on \emph{spatially coupled} sparse graphs has received a lot of attention in the literature. The concept of coupling a sequence of identical small structured graphs into a chain with improved properties, first demonstrated in the context of LDPC convolutional codes~\cite{fz99}, has been shown to be applicable to a diverse list of topics, such as compressed sensing \cite{kp10}, multiuser communication \cite{st11}\cite{ttk11}, quantum codes \cite{hkis11}, and models in statistical physics \cite{hmu10}.\symbolfootnote[0]{This work was partially supported by NSF grant CCF08-30650 and the Alberta Innovates Fund.}

Ensembles of spatially coupled LDPC codes (SC-LDPCCs) can be obtained by terminating regular LDPC convolutional code ensembles \cite{lfzc09}. The slight irregularity resulting from the termination of the convolutional codes has been shown to lead to substantially better belief propagation (BP) decoding thresholds compared to corresponding block, or \emph{uncoupled}, code ensembles for a variety of channels \cite{lscz10,lfzc09,lmfc10b,kmru10}. Further, it has recently been proven analytically for the binary erasure channel (BEC) that the BP decoding thresholds of regular SC-LDPCC ensembles approach the maximum a posteriori probability (MAP) decoding thresholds of the corresponding LDPC block code ensembles \cite{kru11}.


The reduced check node degrees at the ends of the spatially coupled chain result in highly reliable information, which then propagates through the SC-LDPCC chain during the iterative decoding process. As a result, this `flow' of reliable information leads to improved iterative decoding performance. However, a large number of iterations, and consequently a large amount of computation, is needed for the coded symbols in the middle of the chain to be decoded reliably. Improved iterative decoding schedules that attempt to reduce the number of computations per bit to practically reasonable values have recently been proposed. These include windowed decoding~\cite{cip+10}, for which updates of the reliability values are performed within a window sliding along the coupled chain, as well as a decoding schedule where updates are only performed for nodes where there is a potential improvement in the reliability value~\cite{lpf11}. These improved schedules guarantee that the number of computations per bit is independent of the length of the chain. However, further improvement is challenging for single, linear, spatially coupled chains.

In this paper, we propose to connect several individual chains of spatially coupled graphs to form more general graphs. The reliability information that propagates through the individual chains during the iterative decoding process may now spread in multiple directions, thus triggering further performance improvement. This decoding improvement can come from check nodes of reduced degree or variable nodes of increased degree, and it can arise simultaneously in several parts of the graph and then combine to facilitate decoding convergence. We demonstrate improved iterative decoding thresholds for the binary erasure channel (BEC) and the additive white Gaussian noise (AWGN) channel and a reduced number of computations per bit required to reach a desired probability of error. Moreover, we show that the connected structures, like the individual chains, are asymptotically good. Although in this paper we limit our study to a construction consisting of two parallel linear coupled chains connected by two bridges, the concept can be easily generalized.


\section{Code Construction}
\label{sec:construction}

\begin{figure}[h]
\setlength{\unitlength}{1mm}
   \begin{picture}(90,35)
  \put(0,0){\includegraphics{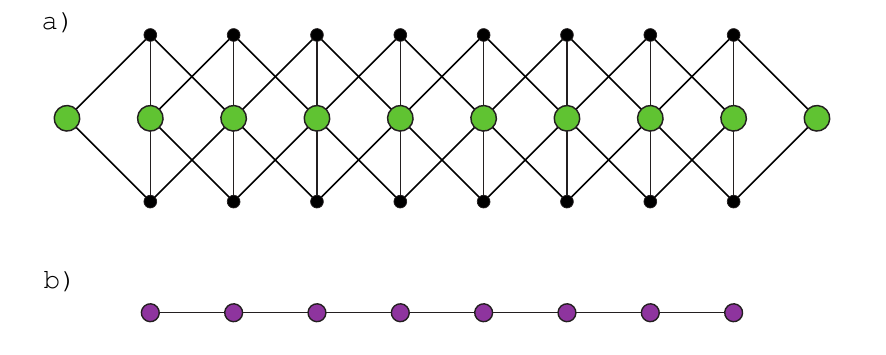}
}
\end{picture}
\caption{A spatially coupled $(3,6)$ protograph chain of length $L=8$ (a) and its simplified representation (b).}
\label{Fig:parallel}
\end{figure}

We start by describing a SC-LDPCC ensemble in terms of its protograph representation. A protograph \cite{tho03} is a small bipartite graph connecting two sets of nodes, a set of variable nodes and a set of check nodes. A protograph which represents a terminated chain of coupled $(3,6)$-regular LDPC codes of length $L=8$ is shown in Fig.~\ref{Fig:parallel}(a). The variable nodes are black and the check nodes are green. Each variable node is connected to $3$ check nodes and each check node is connected to $6$, $4$, or $2$ variable nodes. The associated bi-adjacency matrix $\mathbf{B}$ of this protograph is called the \emph{base} matrix.


The parity-check matrix $\mathbf{H}$ of a protograph-based LDPC block code  (a member of the ensemble) can be created by replacing each non-zero entry in $\mathbf{B}$  by 
a permutation matrix of size $M$ (for protographs with single edge connections) and a zero entry by the $M\times M$ all-zero matrix. In graphical terms, this can be viewed as taking an $M$-fold graph cover 
or ``lifting'' of the protograph. The parameter $M$ is referred to as the lifting factor. It is an important feature of this construction that each lifted code inherits the degree distribution and local graph neighborhood structure of the protograph. 

The protograph chain shown in Fig.~\ref{Fig:parallel}(a) can be viewed as a row of simple vertical graphs (segments) where two variable nodes are connected to the check node located in between. These segments are then coupled, i.e., each segment is connected to one segment on the left and one on the right (see Fig.~\ref{Fig:parallel}(b)). To effect termination, one extra check node is required at each end of the chain. Coupling a sequence of graphs in such a way has two immediate consequences: 1) there is a loss in rate that diminishes as the length $L$ of the chain grows, and 2) we observe a threshold improvement compared to the uncoupled LDPC code ensembles.


\subsection{Construction: Two Chains Connected by Two Bridges}

\begin{figure}[h]
\setlength{\unitlength}{1mm}
   \begin{picture}(90,60)
  \put(0,0){\includegraphics{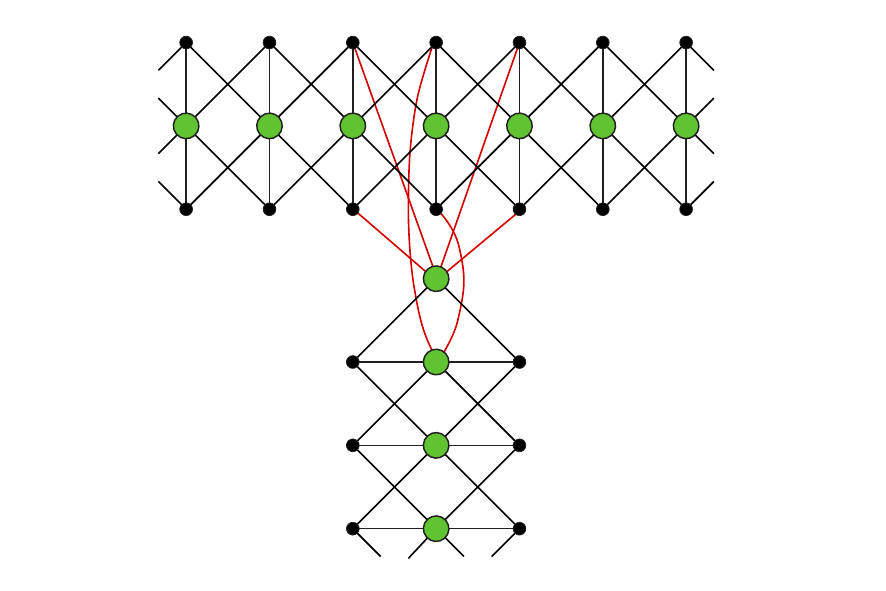}
   }
\end{picture}
\caption{Two connected spatially coupled $(3,6)$-regular protograph chains. The connecting edges are shown in red. 
}
\label{Fig:connector}
\end{figure}

Two protograph chains may be connected as shown in Fig.~\ref{Fig:connector}. 
The horizontal chain continues to the left and right while the vertical chain is terminated at the top. The first check node at the top of the vertical chain has degree two and the second node degree four. Therefore, in order to form degree six check nodes, we connect the first check node of the vertical chain to variable nodes of the horizontal chain with four additional edges and the second check node of the vertical chain to variable nodes of the horizontal chain with two additional edges. As a result, the degrees of several variable nodes in the horizontal chain are increased by one. The new edges are shown in red in Fig.~\ref{Fig:connector}. 

\begin{figure}[h]
\setlength{\unitlength}{1mm}
   \begin{picture}(90,35)
   \put(0,0){\includegraphics{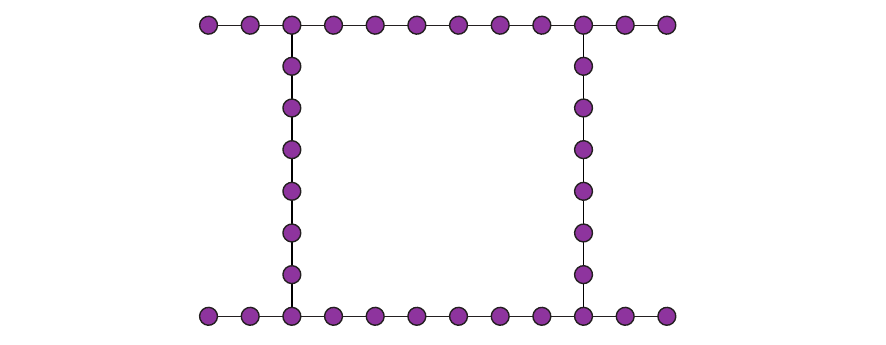}
   }
\end{picture}
\caption{Two parallel $(3,6)$ protograph chains of length $L=12$ connected by two bridges of length $L\stxt{b}=6$ each.}
\label{Fig:2c2b}
\end{figure}

We now consider two horizontal chains connected by two vertical chains that we call bridges. An example of such a construction is given in Fig.~\ref{Fig:2c2b}. The length of both horizontal chains is $12$, while each bridge is of length $6$. Each of the four connections between bridges and
chains is made as shown in Fig.~\ref{Fig:connector}.

Generalizing this example, we consider an ensemble of protographs, denoted by $\mathcal{S}(3,6,L)$, that consists of two chains of length $L$ that are connected by two bridges of length $L/2$. The connecting points are located at a distance of $\lfloor L/4 \rfloor$ from the ends of the chains. The rate of the ensemble $\mathcal{S}(3,6,L)$ is given by
\begin{equation}
R(\mathcal{S}(3,6,L)) = \frac{3L-8}{6L} = \frac{1}{2}-\frac{4}{3L}\ .
\end{equation}


\section{Analysis of connected SC-LDPCCs}
\label{sec:analysis}
\subsection{Iterative decoding analysis}
Since every member of the protograph-based ensemble preserves the structure of the base protograph, density evolution analysis for the resulting codes can be performed within the protograph. In this section, we discuss density evolution for the case of the BEC (iterative decoding threshold results are also presented for the AWGN channel in Section~\ref{sec:results}). We assume that BP decoding is performed after transmission over a BEC with erasure probability $\varepsilon$. In every decoding iteration, all of the check nodes are updated followed by all of the variable nodes. The messages that are passed between the nodes represent either an erasure or the correct symbol value ($0$ or $1$).

We denote the set of variable nodes connected to check node $k$ by $\mathbb{V}(k)$, $k=1,2,\ldots,3L+8$, and the set of check nodes connected to variable node $j$ by $\mathbb{C}(j)$, $j=1,2,\ldots,6L$. The probability that the message passed from check node $k$ to variable node $j$ in iteration $i$ is an erasure is denoted by $q_{kj}^{(i)}$. The probability of an erasure message from variable node $j$ to check node $k$ is similarly denoted by $p_{jk}^{(i)}$. The following equations relate the erasure probabilities of the messages at different iterations:
\begin{align}
q_{kj}^{(i)} &= 1 - \prod_{j' \in \mathbb{V}(k) \smallsetminus j} (1 - p_{j'k}^{(i-1)})\ , \\
p_{jk}^{(i)} &= \epsilon \prod_{k' \in \mathbb{C}(j) \smallsetminus k} q_{k'j}^{(i)}\ .
\end{align}
The variable node messages are initialized as $p_{jk}^{(0)}=\epsilon$ at iteration $0$. The error probability of the variable nodes at iteration $i$ can be calculated as
\begin{equation}
P\stxt{b}(j) = \epsilon \prod_{k \in \mathbb{C}(j) } q_{kj}^{(i)}\ .
\end{equation}

The error probability evolution equations given above indicate that an increase in the number of variable node connections will improve the decoding performance (lower the erasure probability), while an increase in the number of check node connections will degrade it. Therefore, each connection between a protograph chain and a bridge in the ensemble $\mathcal{S}(3,6,L)$ is made by connecting some of the variable nodes in the chain to the check nodes with reduced degrees at the end of the bridge (see Fig.~\ref{Fig:connector}). The degrees of the check nodes in the resulting construction do not exceed $6$, while some of the variable nodes involved in the connection have degree $4$ instead of $3$. 

Focusing on a reduction in complexity, we consider simultaneous decoding of the entire code graph, where we employ the updating schedule proposed in~\cite{lpf11}. The algorithm designates a target convergence probability $P\stxt{b,max}$ as well as an update improvement constraint $\theta$. Regular message passing updates are performed for each variable or check node with the exception of the following conditions:
\begin{itemize}
\item no update for variable node $j$ is performed if the error probability $P\stxt{b}(j) < P\stxt{b,max}$;
\item no update for any variable node $j$, or any check node $k$, is performed if all the nodes in $\mathbb{C}(j)$, or $\mathbb{V}(k)$, respectively, were not updated in the previous iteration;
\item no update for variable node $j$ is performed if the potential improvement of the bit error probability is less than $\theta$, i.e.,
\begin{equation}
\frac{P\stxt{b,old}(j) - P\stxt{b,new}(j)}{P\stxt{b,old}(j)} < \theta\ .
\end{equation}
\end{itemize}

It can be observed that, during every iteration, different sets of nodes are active in the graph. The convergence of the error probability spreads from the ends of the parallel chains into the main square of the graph (see Fig.~\ref{Fig:2c2b}). As will become evident later in this section, the sets of nodes with the fastest convergence are known in advance, so scheduling can be arranged such that the best set of nodes is transmitted, received, and decoded first. This set is then followed by the set of nodes with the second fastest convergence, which is then transmitted and decoded, and so on. This approach, suitable for the ensemble $\mathcal{S}(3,6,L)$, can be seen as an attractive alternative to the windowed decoding approach (which is a good choice for decoding an individual SC-LDPCC chain $\mathcal{C}(3,6,L)$).\footnote{In windowed decoding~\cite{{cip+10}}, it is assumed that the decoder operates on a window of $W$ coupled graph segments. All of the graph segments preceding the current window are already decoded, and all of the graph segments following the window are yet to be processed.}

\begin{figure}[h]
\setlength{\unitlength}{1mm}
   \begin{picture}(80,75)
   \put(0,0){\includegraphics{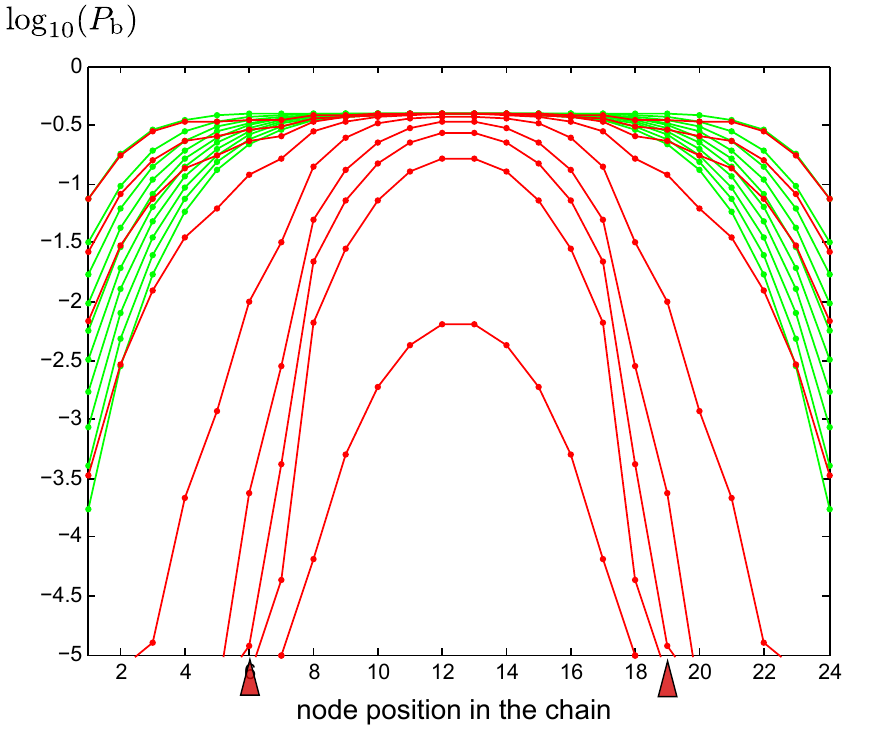}
   }
\end{picture}
\caption{Logarithm of the bit error probability for the variable nodes of the first chain of the ensemble $\mathcal{S}(3,6,24)$ (red curves) and for the ensemble $\mathcal{C}(3,6,24)$ (green curves), as a function of the position of the node in the chain. The curves (either red or green) are computed for decoding  iterations $1,6,11,\ldots,51$ (from top to bottom). The positions of the two bridges are shown by the red triangles.}
\label{Fig:Pb_chain}
\end{figure}

We now shift our focus to the error probability performance of the ensemble $\mathcal{S}(3,6,L)$. We consider SC-LDPCC chains of length $L=24$, a target error probability $P\stxt{b,max}=10^{-5}$, and an update threshold $\theta = 10^{-2}$. The evolution of the bit error probability for the first horizontal chain of the ensemble $\mathcal{S}(3,6,24)$ is depicted in Fig.~\ref{Fig:Pb_chain}. The red curves (from top to bottom) correspond to the error probabilities $P\stxt{b}$ at iterations $i=1,6,11,\ldots,51$. The green curves are the error probabilities $P\stxt{b}$ computed for the single chain ensemble $\mathcal{C}(3,6,24)$ at iterations $i=1,6,11,\ldots,51$. We can see that the variable nodes in the chain that are connected to the two bridges form two clusters which start to converge faster than the remaining nodes in the chain. Although the clusters are not strong enough to make the entire chain converge to the desired $P\stxt{b,max}$, they facilitate the overall convergence process. We observe convergence for the connected SC-LPDCC ensemble $\mathcal{S}(3,6,24)$ happening much faster than that of the ordinary SC-LDPCC ensemble $\mathcal{C}(3,6,24)$.

\begin{figure}[h]
\setlength{\unitlength}{1mm}
   \begin{picture}(80,90)
   \put(0,0){\includegraphics{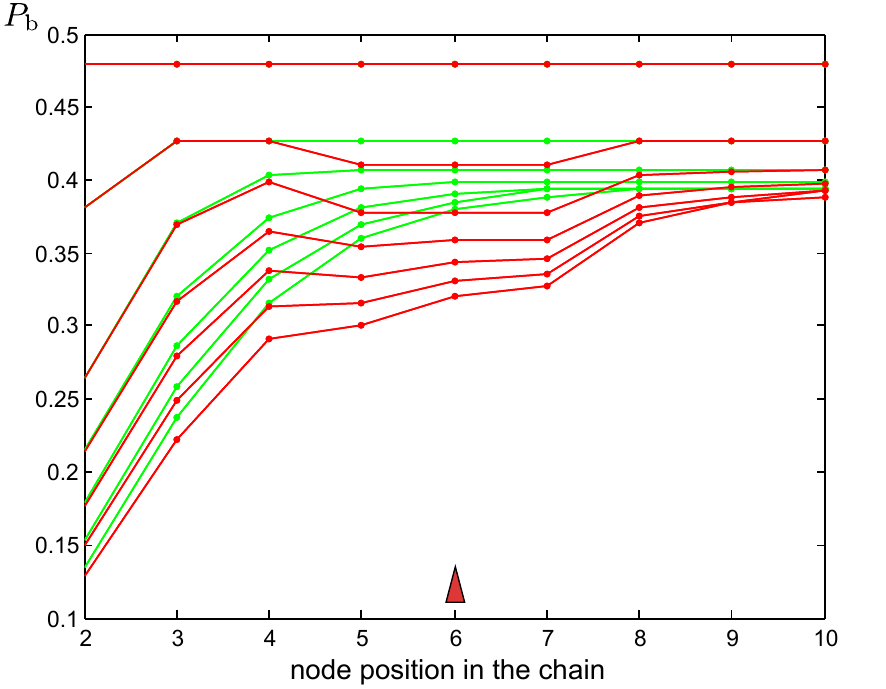}
   }
\end{picture}
\caption{Bit error probability for the variable nodes of the first chain around the bridge connection. The red curves correspond to the ensemble $\mathcal{S}(3,6,24)$ the green curves to the ensemble $\mathcal{C}(3,6,24)$. The curves (either red or green) are computed for decoding  iterations $1,2,\ldots,7$ (from top to bottom). The position of the bridges is shown by the red triangle.}
\label{Fig:Pb_chain_small}
\end{figure}

Fig.~\ref{Fig:Pb_chain_small} presents a magnified picture of the $P\stxt{b}$ evolution. The red curves correspond to the ensemble $\mathcal{S}(3,6,24)$ and the green curves to the ensemble $\mathcal{C}(3,6,24)$. Both sets of curves are computed for iterations $i=1,2,\ldots,7$. We notice that the error probability at the position of the bridge connection becomes irregular compared to its neighborhood, even for the second iteration. As a result, the nodes with higher reliability appear at the four bridge connection positions along the chain. These nodes force the intermediate low reliability nodes to converge faster. Consequently, in the initial iterations, the irregularity of the error probability curve facilitates convergence, and the usual bell-like shape only appears as the iterations proceed. In contrast, the green $P\stxt{b}$ curves, corresponding to the individual SC-LDPCC chain, assume  a smooth bell-like shape from the beginning, and we observe that there is no irregularity present to speed up convergence.

\begin{figure}[h]
\setlength{\unitlength}{1mm}
   \begin{picture}(80,90)
   \put(0,0){\includegraphics{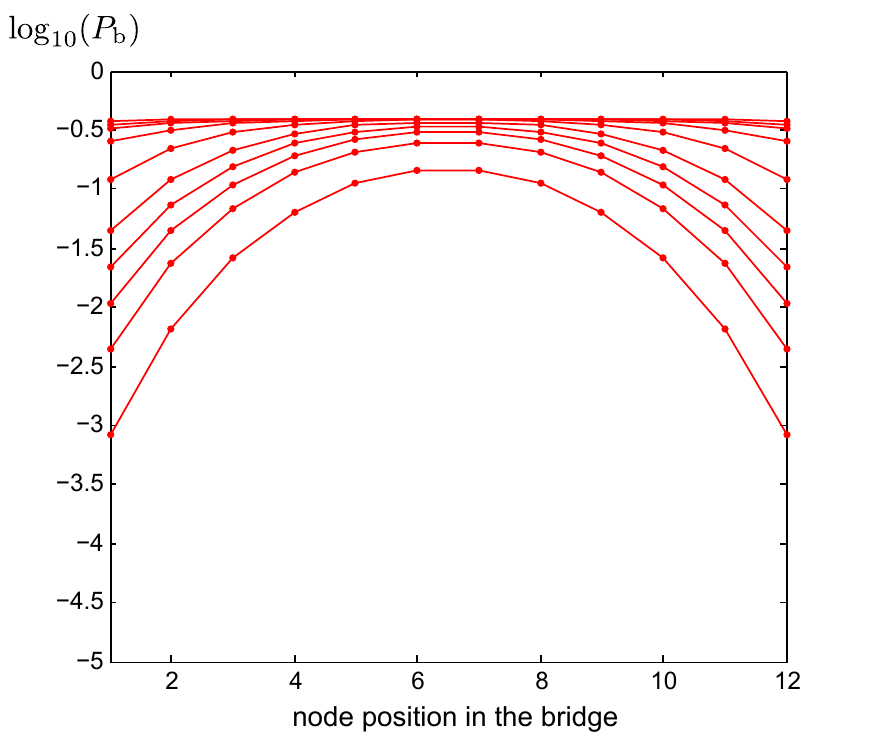}
   }
\end{picture}
\caption{Logarithm of the bit error probability for the variable nodes of the first bridge as a function of the position of the node in the bridge. The curves (from top to bottom) correspond to iterations $1,6,11,\ldots,51$.}
\label{Fig:Pb_bridge}
\end{figure}

The evolution of the probabilities $P\stxt{b}$ for the nodes in the bridge is shown in Fig.~\ref{Fig:Pb_bridge}. The curves correspond to iterations $i=1,6,11,\ldots,51$. Convergence in the bridge starts slowly, since the connections, even though the `strong' variable nodes of higher degree are included, are not able to produce the same boost of reliability information as the low degree check nodes at the ends of each parallel chain. However, as the parallel chains converge, they essentially ``disappear'' and the bridge appears to be left disconnected, or ``hanging in the air''. At that time, the convergence of the bridge evolves as if it were a regular SC-LDPCC chain. Note that the length $L/2$ of the bridges has been selected in order to ensure that their convergence occurs simultaneously with the convergence of the parallel chains.

\subsection{Minimum distance analysis}
In \cite{div06}, Divsalar presented a technique to calculate the average weight enumerator for protograph-based block code ensembles. This weight enumerator can be used to test if an ensemble is \emph{asymptotically good}, i.e., if the minimum distance typical of most members of the ensemble is at least as large as $\delta_{min}n$, where $\delta_{min}$ its the \emph{minimum distance growth rate} of the ensemble and $n$ is the block length. In \cite{lmfc10}, it was shown that ensembles of $(J,K)$-regular SC-LDPCCs (i.e., individual chains) are asymptotically good. In Section~\ref{sec:results}, we present the results of  a similar protograph-based analysis for ensembles of connected SC-LDPCCs to see if they share the good distance properties of the individual chains.

\section{Results}
\label{sec:results}

In this section, we present a number of results for the constructed $\mathcal{S}(3,6,L)$ ensembles. We start by considering transmission over the BEC using the complexity saving decoding schedule discussed in Section~\ref{sec:analysis}. The target error probability $P\stxt{b,max}$ is set to $10^{-5}$. The number of updates per node $I\stxt{eff}$ (for both check and variable nodes, including the chains and bridges), averaged over the node positions, is considered as a measure of decoding complexity.

\begin{figure}[h]
\setlength{\unitlength}{1mm}
   \begin{picture}(90,75)
   \put(0,0){\includegraphics{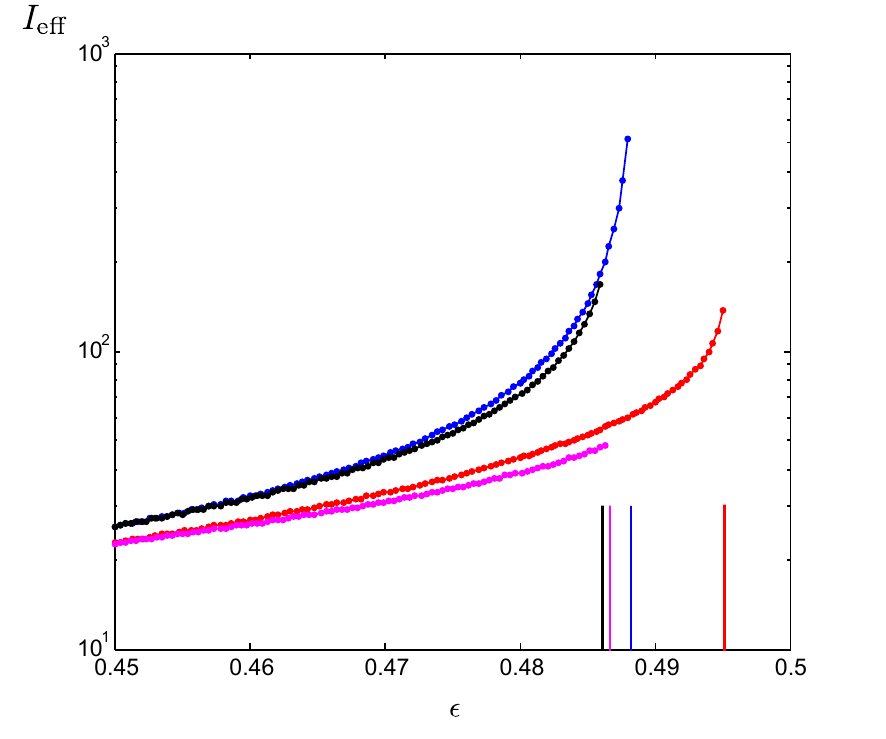}
   }
\end{picture}
\caption{The average number of updates per node $I\stxt{eff}$ as a function of the BEC parameter $\epsilon$ for the $\mathcal{S}(3,6,24)$ ensemble (magenta and red curves) and the $\mathcal{C}(3,6,18)$ ensemble (blue and black curves). The magenta and black curves are computed for the updating schedule with improvement constraint $\theta=10^{-2}$, while the red and blue curve are for $\theta=0$. Corresponding thresholds are given by vertical lines.}
\label{Fig:Ieff}
\end{figure}

The average number of updates per node $I\stxt{eff}$ required to achieve $P\stxt{b,max}=10^{-5}$ is plotted in Fig.~\ref{Fig:Ieff} as a function of the BEC parameter $\epsilon$. The red and magenta curves correspond to the proposed ensemble $\mathcal{S}(3,6,24)$, while the blue and black curves are computed for a single SC-LDPCC chain of length $L=18$, i.e., ensemble $\mathcal{C}(3,6,18)$.\footnote{Ensembles $\mathcal{S}(3,6,24)$ and $\mathcal{C}(3,6,18)$ have been selected for comparison since they both have design rates approximately equal to $0.444$.}. The magenta and black curves correspond to the updating schedule with the improvement constraint $\theta=10^{-2}$, while the red and blue curves correspond to $\theta=0$ (in which case, updates are performed regardless of the potential error probability improvement).  We  observe a significant complexity improvement provided by the connected SC-LDPCC chain construction.\footnote{Although 12 variable nodes in the ensemble $\mathcal{S}(3,6,24)$ have degree 4 instead of 3, the difference in the computation per node it causes is insignificant with regard to the overall obtained improvement.}  The vertical straight lines indicate the iterative decoding thresholds calculated for each construction with the corresponding update schedule, and we note that the ensemble $\mathcal{S}(3,6,24)$ has better iterative decoding thresholds then $\mathcal{C}(3,6,18)$. 

Iterative decoding thresholds for a number of $\mathcal{S}(3,6,L)$ ensembles on the BEC are given in Table~\ref{tab:thres}. The thresholds are compared with the thresholds of the regular coupled chains of (approximately) the same rate. The thresholds for the AWGN channel, computed using a discretized density evolution method, are given in Table~\ref{tab:thresAWGN}. An improvement in the threshold values can be observed for all SC-LDPCC chain lengths $L=8,12,16,20$, and $24$ considered.
\begin{table}[h]
\begin{center}
\scalebox{0.9}{%
\begin{tabular}{|c|c|c|c|c|c|}\hline
Rate & Ensemble & $\epsilon^*$ & Ensemble & $\epsilon^*$ \\\hline
$0.3333$ & $\mathcal{S}(3,6,8)$ &  $0.563$ & $\mathcal{C}(3,6,6)$ & $0.557$\\
$0.3889$ & $\mathcal{S}(3,6,12)$ & $0.538$  & $\mathcal{C}(3,6,9)$ & $0.512$\\
$0.4167$ & $\mathcal{S}(3,6,16)$ & $0.522$ & $\mathcal{C}(3,6,12)$ & $0.495$\\
$0.4333$ & $\mathcal{S}(3,6,20)$ &  $0.504$ & $\mathcal{C}(3,6,15)$ & $0.489$\\
$0.4444$ & $\mathcal{S}(3,6,24)$ &  $0.495$ & $\mathcal{C}(3,6,18)$ &  $0.488$ \\
\hline
\end{tabular}}
\end{center}
\caption{BEC thresholds $\epsilon^*$ for several SC-LDPCC ensembles $\mathcal{S}(3,6,L)$ and single chain ensembles.}\label{tab:thres}\vspace{-4mm}
\end{table}

\begin{table}[h]
\begin{center}
\scalebox{0.9}{%
\begin{tabular}{|c|c|c|c|c|c|}\hline
Rate & Ensemble & $(E\stxt{b}/N_0)^*$ & Ensemble & $(E\stxt{b}/N_0)^*$ \\\hline
$0.3333$ & $\mathcal{S}(3,6,8)$ &  $1.0167$dB & $\mathcal{C}(3,6,6)$ & $1.1894$dB\\
$0.3889$ & $\mathcal{S}(3,6,12)$ & $0.7512$dB  & $\mathcal{C}(3,6,9)$ & $1.1701$dB\\
$0.4167$ & $\mathcal{S}(3,6,16)$ & $0.7231$dB & $\mathcal{C}(3,6,12)$ & $1.1167$dB\\
$0.4333$ & $\mathcal{S}(3,6,20)$ &  $0.8079$dB & $\mathcal{C}(3,6,15)$ & $1.0431$dB\\
$0.4444$ & $\mathcal{S}(3,6,24)$ &  $0.8367$dB & $\mathcal{C}(3,6,18)$ & $0.9659$dB\\
\hline
\end{tabular}}
\end{center}
\caption{AWGN channel thresholds $(E\stxt{b}/N_0)^*$ for several SC-LDPCC ensembles $\mathcal{S}(3,6,L)$ and single chain ensembles.}\label{tab:thresAWGN}\vspace{-4mm}
\end{table}

Minimum distance growth rates for several connected $\mathcal{S}(3,6,L)$ SC-LDPCC ensembles are shown in Table~\ref{tab:dist}. We observe that, like the individual component SC-LDPCC chains, the $\mathcal{S}(3,6,L)$ ensembles are asymptotically good. As the length of the parallel SC-LDPCC chains increases, the design rate increases, the thresholds approach capacity, and the minimum distance growth rate decreases. This is analogous to the effect of increasing the length $L$ of the individual SC-LDPC chain $\mathcal{C}(3,6,L)$ \cite{lmfc10}. However, the main advantage that a connected SC-LDPCC ensemble $\mathcal{S}(3,6,L)$ has over the individual ensemble $\mathcal{C}(3,6,L)$ is that the connected structure of $\mathcal{S}(3,6,L)$ permits the reliability information to spread in multiple directions and facilitate decoding convergence, resulting in a reduced number of computations per bit required to reach a desired probability of error in addition to a threshold improvement. 



\begin{table}[h]
\begin{center}
\scalebox{0.9}{%
\begin{tabular}{|c|c|c||c|c|c|}\hline
$L$ & Rate & $\delta_{min}$ & $L$ & Rate & $\delta_{min}$\\\hline
$8$ & $1/3$ &  $0.0137$ & $16$ & $5/12$ & $0.0054$ \\
$10$ & $11/30 $ &$0.0095$ & $18$ & $23/54$ & $0.0049$ \\
$12$ & $7/18$ & $0.0075$ & $20$ & $13/30$ & $0.0043$ \\
$14$ & $17/42$ & $0.0062$ & $24$ & $4/9$ & $0.0036$ \\
\hline
\end{tabular}}
\end{center}
\caption{Minimum distance growth rates for several connected SC-LDPCC ensembles
$\mathcal{S}(3,6,L)$.}\label{tab:dist}\vspace{-5mm}
\end{table}

%
%
%

\section{Conclusions}
\label{sec:conc}

In this paper, a construction of asymptotically good protograph-based ensembles using connected SC-LDPCC chains was presented. Transmission over both the BEC and the AWGN channel along with iterative message passing decoding have been considered. It was demonstrated that connecting coupled chains improves the iterative decoding thresholds and reduces the decoding complexity per bit compared to individual SC-LDPC chains. The results suggest that the threshold saturation effect shown in \cite{kru11} may be even more pronounced when the coupling is not restricted to a single linear connection.

\end{document}